\documentclass[preprint]{revtex4-1}

\usepackage{graphicx,amsmath,amssymb}
\usepackage{amsfonts}

\begin{document}
\title{Weak value amplification: a view from quantum estimation theory that highlights what it is and what isn't}

\author{Juan P. Torres}
\email{Corresponding author: juanp.torres@icfo.es}
\affiliation{ICFO-Institut de Ciencies Fotoniques, Mediterranean Technology Park, 08860 Castelldefels (Barcelona), Spain
}%
\affiliation{Dep. Signal Theory and Communications, Universitat Polit{\`{e}}cnica de Catalunya, 08034 Barcelona, Spain
}%

\author{Luis Jos\'{e} Salazar-Serrano}%
\affiliation{ICFO-Institut de Ciencies Fotoniques, Mediterranean Technology Park, 08860 Castelldefels (Barcelona), Spain
}%
\affiliation{Quantum Optics Laboratory, Universidad de los Andes, AA 4976, Bogot\'{a}, Colombia
}%

\date{\today}

\begin{abstract}
Weak value amplification (WVA) is a concept that has been
extensively used in a myriad of applications with the aim of
rendering measurable tiny changes of a variable of interest. In
spite of this, there is still an on-going debate about its {\em
true} nature and whether is really needed for achieving high
sensitivity. Here we aim at solving the puzzle, using some basic
concepts from quantum estimation theory, highlighting what the use
of the WVA concept can offer and what it can not. While WVA cannot
be used to go beyond some fundamental sensitivity limits that
arise from considering the full nature of the quantum states, WVA
can notwithstanding enhance the sensitivity of {\em real}
detection schemes that are limited by many other things apart from
the quantum nature of the states involved, i.e. {\em technical
noise}. Importantly, it can do that in a straightforward and
easily accessible manner.
\end{abstract}

\maketitle

\section*{Introduction}
Weak value amplification (WVA) \cite{aharonov1988} is a concept
that has been used under a great variety of experimental
conditions
\cite{hosten2008,zhou2012,ben_dixon2009,pfeifer2011,howell2010_freq,egan2012,xu_guo2013}
to reveal tiny changes of a variable of interest. In all those
cases, a priori sensitivity limits were not due to the quantum
nature of the light used ({\em photon statistics}), but instead to
the insufficient resolution of the detection system, what might be
termed generally as {\em technical noise}. WVA was a feasible
choice to go beyond this limitation. In spite of this extensive
evidence, ``its interpretation has historically been a subject of
confusion" \cite{dressel2014}. For instance, while some authors
\cite{jordan2014} show that ``weak-value-amplification techniques
(which only use a small fraction of the photons) compare favorably
with standard techniques (which use all of them)", others
\cite{knee2014} claim that WVA ``does not offer any fundamental
metrological advantage" , or that WVA \cite{ferrie2014} ``does not
perform better than standard statistical techniques for the tasks
of single parameter estimation and signal detection''. However,
these conclusions are criticized by others based on the idea that
``the assumptions in their statistical analysis are irrelevant for
realistic experimental situations'' \cite{vaidman2014}. The
problem might reside in

Here we make use of some simple, but fundamental, results from
quantum estimation theory \cite{helstrom1976} to show that there
are two sides to consider when analyzing in which sense WVA can be
useful. On the one hand, the technique generally makes use of
linear-optics unitary operations. Therefore, it cannot modify the
statistics of photons involved. Basic quantum estimation theory
states that the post-selection of an appropriate output state, the
basic element in WVA, cannot be better than the use of the input
state \cite{nielsen2000}. Moreover, WVA uses some selected,
appropriate but partial, information about the quantum state that
cannot be better that considering the full state. Indeed, due to
the unitarian nature of the operations involved, it should be
equally good any transformation of the input state than performing
no transformation at all. In other words, when considering only
the quantum nature of the light used, WVA cannot enhance the
precision of measurements \cite{lijian2015}.

On the other hand, a more general analysis that goes beyond only
considering the quantum nature of the light, shows that WVA can be
useful when certain technical limitations are considered. In this
sense, it might increase the ultimate resolution of the detection
system by effectively lowering the value of the smallest quantity
that can detected. In most scenarios, although not always
\cite{torres2012}, the signal detected is severely depleted, due
to the quasi-orthogonality of the input and output states
selected. However, in many applications, limitations are not
related to the low intensity of the signal \cite{hosten2008}, but
to the smallest change that the detector can measure
irrespectively of the intensity level of the signal.

A potential advantage of our approach is that we make use of the
concept of trace distance, a clear and direct measure of the
degree of distinguishability of two quantum states. Indeed, the
trace distance gives us the minimum probability of error of
distinguishing two quantum states that can be achieved under the
best detection system one can imagine \cite{helstrom1976}.
Measuring tiny quantities is essentially equivalent to
distinguishing between nearly parallel quantum states. Therefore
we offer a very basic and physical understanding of how WVA works,
based on the idea of how WVA transforms very close quantum states,
which can be useful to the general physics reader.

Here were we use an approach slightly different from what other
analysis of WVA do, where most of the times the tool used to
estimate its usefulness is the Fisher information. Contrary to how
we use the trace distance here, to set a sensitivity bound only
considering how the quantum state changes for different values of
the variable of interest, the Fisher information requires to know
the probability distribution of possible experimental outcomes for
a given value of the variable of interest. Therefore, it can look
for sensitivity bounds for measurements by including {\em
technical characteristics} of specific detection schemes
\cite{jordan2014}. A brief comparison between both approaches will
be done towards the end of this paper.

One word of caution will be useful here. The concept of weak value
amplification is presented for the most part in the framework of
Quantum Mechanics theory, where it was born. It can be readily
understood in terms of constructive and destructive interference
between probability amplitudes \cite{duck1989}. Interference is a
fundamental concept in any theory based on waves, such as
classical electromagnetism. Therefore, the concept of weak value
amplification can also be described in many scenarios in terms of
interference of classical waves \cite{howell2010}. Indeed, most of
the experimental implementations of the concept, since its first
demonstration in 1991 \cite{ritchie1991}, belong to this type  and
can be understood without resorting to a quantum theory formalism.

\subsection*{An example of the application of the weak value amplification concept: measuring small temporal delays with large bandwidth pulses.}
For the sake of example, we consider a specific weak amplification
scheme \cite{brunner2010}, depicted in Fig. 1,
which has been recently demonstrated experimentally
\cite{xu_guo2013,salazar2014}. It aims at measuring very small
temporal delays $\tau$, or correspondingly tiny phase changes
\cite{strubi2013}, with the help of optical pulses of much larger
duration. We consider this specific case because it contains the
main ingredients of a typical WVA scheme, explained below, and it
allows to derive analytical expressions of all quantities
involved, which facilitates the analysis of main results.
Moreover, the scheme makes use of linear optics elements only and
also works with large-bandwidth partially-coherent light
\cite{li_guo2011}.

In general, a WVA scheme requires three main ingredients: a) the
consideration of two subsystems (here two degrees of freedom: the
polarisation and the spectrum of an optical pulse) that are weakly
coupled (here we make use of a polarisation-dependent temporal
delay that is introduced with the help of a Michelson
interferometer); b) the {\em pre-selection} of the input state of
both subsystems; and c) the {\em post-selection} of the state in
one of the subsystems (the state of polarisation) and the
measurement of the state of the remaining subsystem (the spectrum
of the pulse). With appropriate {\em pre-} and {\em
post-selection} of the polarisation of the output light, tiny
changes of the temporal delay $\tau$ can cause anomalously large
changes of its spectrum, rendering in principle detectable very
small temporal delays.

\begin{figure}
\centering
\includegraphics[width=0.7\textwidth]{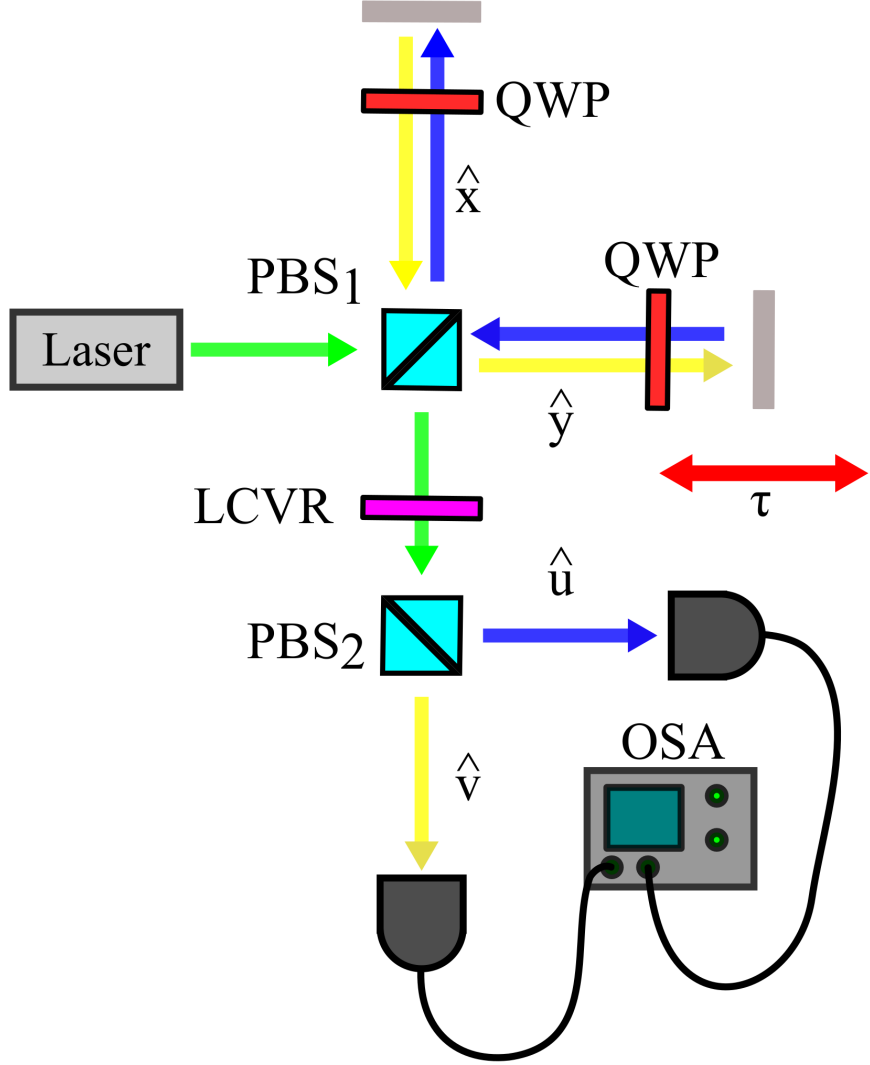}
\caption{Weak value amplification scheme aimed at detecting
extremely small temporal delays. The input pulse polarisation
state is selected to be left-circular by using a polariser, a
quarter-wave plate (QWP) and a half-wave plate (HWP). A first
polarising beam splitter (PBS$_1$) splits the input into two
orthogonal linear polarisations that propagate along different
arms of the interferometer. An additional QWP is introduced in
each arm to rotate the beam polarisation by $90^{\circ}$ to allow
the recombination of both beams, delayed by a temporal delay
$\tau$, in a single beam by the same PBS. After PBS$_1$, the
output polarisation state is selected with a liquid crystal
variable retarder (LCVR) followed by a second polarising beam
splitter (PBS$_2$). The variable retarder is used to set the
parameter $\theta$ experimentally. Finally, the spectrum of each
output beam is measured using an optical spectrum analyzer (OSA).
($\hat{x}$,$\hat{y}$) and ($\hat{u}$,$\hat{v}$) correspond to two
sets of orthogonal polarisations. Figure drawn by one of the
authors (Luis-Jose Salazar Serrano).} \label{figure_scheme}
\end{figure}

Let us be more specific about how all these ingredients are
realized in the scheme depicted in Fig. 1. An
input coherent laser beam ($N$ photons) shows circular
polarisation, ${\bf e}_{\mathrm{in}}=1/\sqrt{2}\,\left(
\hat{x}-i\hat{y} \right)$, and a Gaussian shape with temporal
width $T_0$ (Full-width-half maximum, $\tau \ll T_0$). The
normalized temporal and spectral shapes of the pulse read
\begin{eqnarray}
& & \Psi(t)=\left( \frac{4 \ln 2}{\pi T_0^2}\right)^{1/4} \exp
\left(
-\frac{2 \ln 2 t^2}{T_0^2} \right) \nonumber \\
& & \Psi(f)=\left(\frac{\pi T_0^2}{\ln 2}\right)^{1/4} \exp \left(
-\frac{\pi^2 T_0^2 f^2}{2 \ln 2} \right).
\end{eqnarray}
The input beam is divided into the two arms of a Michelson
interferometer with the help of a polarising beam splitter
(PBS$_1$). Light beams with orthogonal polarisations traversing
each arm of the interferometer are delayed $\tau_0$ and
$\tau_0+\tau$, respectively, which constitute the weak coupling
between the two degrees of freedom. After recombination of the two
orthogonal signals in the same PBS$_1$, the combination of a
liquid-crystal variable retarder (LCVR) and a second polarising
beam splitter (PBS$_2$) performs the post-selection of the
polarisation of the output state, projecting the incoming signal
into the polarisation states $\hat{u}=1/\sqrt{2} \left[
\hat{x}+\hat{y}\exp(i\theta) \right]$ and $\hat{v}=1/\sqrt{2}
\left[ \hat{x}-\hat{y}\exp(i\theta) \right]$. The amplitudes of
the signals in the two output ports write (not normalized)
\begin{eqnarray} & &  \Phi_u(\tau)=\frac{\Psi(\Omega)}{2} \exp
\left[ i(\omega_0 +\Omega) \tau_0 \right] \left\{1+
\exp \left[ i (\omega_0+\Omega) \tau-i\Gamma \right] \right\} \label{projections1} \\
& & \Phi_v(\tau)=\frac{\Psi(\Omega)}{2} \exp \left[ i \left(
\omega_0+ \Omega \right)\tau_0 \right] \left\{ 1-\exp \left[ i
(\omega_0+\Omega) \tau -i\Gamma \right] \right\},
\label{projections2}
\end{eqnarray}
where $\Gamma=\pi/2+\theta$.

\begin{figure}
\centering
\includegraphics[width=0.9\textwidth]{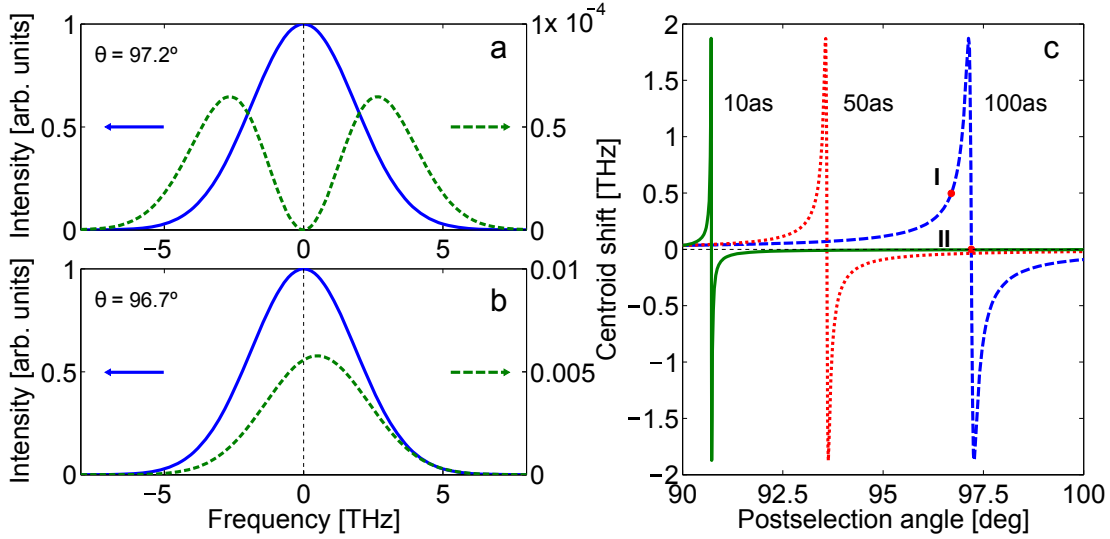}
\caption{Spectrum measured at the output. (a) and (b): Spectral shape of the mode functions for $\tau=0$  (solid blue line) and $\tau=100$ as (dashed green line). In (a) the post-selection angle $\theta$ is $97.2^{\circ}$, so as to fulfil the condition $\omega_0 \tau-\Gamma=\pi$. In (b) the angle $\theta$ is $96.7^{\circ}$. (c) Shift of the centroid of the spectrum of the output pulse after projection into the polarisation state $\hat{u}$ in PBS$_2$, as a function of the post-selection angle $\theta$. Green solid line: $\tau=10$ as; Dotted red line: $\tau=50$ as, and dashed blue line: $\tau=100$ as. Label {\bf I} corresponds to $\theta=96.7^{\circ}$ [mode for $\tau=100$ as shown in (b)]. Label {\bf II} corresponds to $\theta=97.2^{\circ}$, where the condition $\omega_0 \tau-\Gamma=\pi$ is fulfiled [mode for $\tau=100$ shown in (a)]. It yields the minimum mode overlap between states with $\tau=0$ and $\tau \neq 0$. Data: $\lambda_0=1.5\, \mu$m and $T_0=100$ fs.} \label{figure_modes}
\end{figure}

After the signal projection performed after PBS$_2$, the WVA
scheme distinguishes different states, corresponding to different
values of the temporal delay $\tau$, by measuring the spectrum of
the outgoing signal in the selected output port.  The different
spectra obtained for delays $\tau=0$ and $\tau=100$ as, for two
different polarisation projections, are shown in Figures 2 (a) and
2 (b). To characterize different modes one can measure, for
instance, the centroid of the spectrum. Fig. 2 (c) shows the
centroid shift of the output signal for $\tau \ne 0$, which reads
\begin{equation} \Delta
f=-\frac{\tau\,\ln 2 }{\pi T_0^2}  \frac{\gamma \sin
\left(\omega_0 \tau - \Gamma\right)}{1+\gamma \cos \left(\omega_0
\tau-\Gamma\right)}, \label{centroid_shift}
\end{equation}
The differential power between both signals (with $\tau=0$ and
$\tau \ne 0$) reads
\begin{equation}
\frac{P_{\mathrm{out}}(\tau)-P_{\mathrm{out}}(\tau=0)}{P_{\mathrm{in}}}=\frac{1}{2}\,
\left[ \cos \Gamma-\cos \left(\omega_0 \tau -\Gamma \right)
\right]
\end{equation}
When there is no polarisation-dependent time delay ($\tau=0$), the
centroid of the spectrum of the output signal is the same than the
centroid of the input laser beam, i.e., there is no shift of the
centroid ($\Delta f=0$). However, the presence of a small $\tau$
can produce a large and measurable shift of the centroid of the
spectrum of the signal.

\section*{Results}
\subsection*{View of weak value amplification from quantum estimation theory}
Detecting the presence ($\tau \neq 0$) or absence ($\tau=0$) of a
temporal delay between the two coherent orthogonally-polarised
beams after recombination in PBS$_1$, but before traversing
PBS$_2$, is equivalent to detecting which of the two quantum
states,
\begin{equation}
|\Phi_0 \rangle=|\Phi (\tau_0) \rangle_x |\Phi (\tau_0) \rangle_y
\end{equation} or
\begin{equation}
|\Phi_1 \rangle=|\Phi(\tau_0) \rangle_x |\Phi (\tau_0+\tau)
\rangle_y \label{state1}
\end{equation}
is the output quantum state which describes the coherent pulse
leaving PBS$_1$. $(x,y)$ designates the corresponding
polarisations. The spectral shape (mode function) $\Phi$ writes
\begin{equation}
\Phi(\tau_0+\tau) = \Psi(\Omega) \exp \left[ i (\omega_0+\Omega)
(\tau_0+\tau) \right],  \label{modes_input}
\end{equation}
where $\omega_0$ is the central frequency of the laser pulse,
$\Omega=2\pi f$ is the angular frequency deviation from the the
center frequency and $\Psi(\Omega)$ is the spectral shape of the
input coherent laser signal.

The minimum probability of error that can be made when
distinguishing between two quantum states is related to the trace
distance between the states \cite{fuchs1999}. For two pure state,
$\Phi_0$ and $\Phi_1$, the (minimum) probability of error is
\cite{helstrom1976,ou1996,englert1996}
\begin{equation}
\label{average_error}
P_{\mathrm{error}}=\frac{1}{2}\left(1-\sqrt{1-|\langle
\Phi_0|\Phi_1\rangle|^2} \right).
\end{equation}
For $\Phi_0=\Phi_1$, $P_{\mathrm{error}}=0.5$. On the contrary, to
be successful in distinguishing two quantum states with low
probability of error ($P_{\mathrm{error}} \sim 0$) requires
$|\langle \Phi_0|\Phi_1\rangle| \sim 0$, i.e., the two states
should be close to orthogonal.

The coherent broadband states considered here can be generally
described as single-mode quantum states where the mode is the
corresponding spectral shape of the light pulse. Let us consider
two single-mode coherent beams
\begin{eqnarray}
& & |\alpha \rangle=\exp \left( -\frac{|\alpha|^2}{2} \right)
\sum_{n=0}^{\infty} \frac{\alpha^n \left(
A^{\dagger}\right)^n}{n!}|0\rangle \nonumber \\
& & |\beta \rangle=\exp \left( -\frac{|\beta|^2}{2} \right)
\sum_{n=0}^{\infty} \frac{\beta^n \left(
B^{\dagger}\right)^n}{n!}|0\rangle,
\end{eqnarray}
where $A$ and $B$ are the two modes
\begin{eqnarray}
& & A^{\dagger}=\int d\Omega F(\Omega) a^{\dagger}(\Omega) \nonumber \\
& & B^{\dagger}=\int d\Omega G(\Omega) a^{\dagger}(\Omega),
\end{eqnarray}
and $|\alpha|^2$ and $|\beta|^2$ are the mean number of photons in
modes $A$ and $B$, respectively. The mode functions $F$ and $G$
are assumed to be normalized, i.e., $\int d\Omega
|F(\Omega)|^2=\int d\Omega |G(\Omega)|^2=1$. The overlap between
the quantum states, $|\langle \beta|\alpha \rangle|^2$, reads
\begin{equation}
\label{overlap1}|\langle \beta|\alpha \rangle|^2=\exp \left(
-|\alpha|^2-|\beta|^2 + \rho \alpha \beta^{*}+\rho^{*}\alpha^{*}
\beta\right),
\end{equation}
where we introduce the mode overlap $\rho$ that reads
\begin{equation}
\label{overlap2} \rho=\int d\Omega F(\Omega) \left[ G(\Omega)
\right]^{*}.
\end{equation}
In order to obtain Eq. (\ref{overlap1}) we have made use of
$\langle 0|B^n \left[ A^{\dagger}\right]^m |0 \rangle=n! \rho^n
\delta_{nm}$. For $\rho=1$ (coherent beams in the same mode but
with possibly different mean photon numbers) we recover the
well-known formula for single-mode coherent beams
\cite{glauber1963}: $|\langle \beta|\alpha
\rangle|^2=\exp\left(-|\alpha-\beta|^2 \right)$.

Making use of Eqs. (\ref{modes_input}), (\ref{overlap1}) and
(\ref{overlap2}) we obtain
\begin{eqnarray}
& & |\langle \Phi_0|\Phi_1\rangle|^2 \nonumber \\
& & =|\langle \Phi(\tau_0)|\Phi(\tau_0)\rangle_x|^2  |\langle
\Phi(\tau_0)|\Phi(\tau_0+\tau)\rangle_y|^2  \nonumber \\
& & = \exp \left[-N \left( 1-\gamma \cos \omega_0 \tau\right)
\right], \label{input_result}
\end{eqnarray}
where
\begin{equation}
\gamma= \exp \left( -\ln 2 \,\frac{\tau^2}{T_0^2} \right).
\end{equation}

\begin{figure}
\centering
\includegraphics[width=0.5\textwidth]{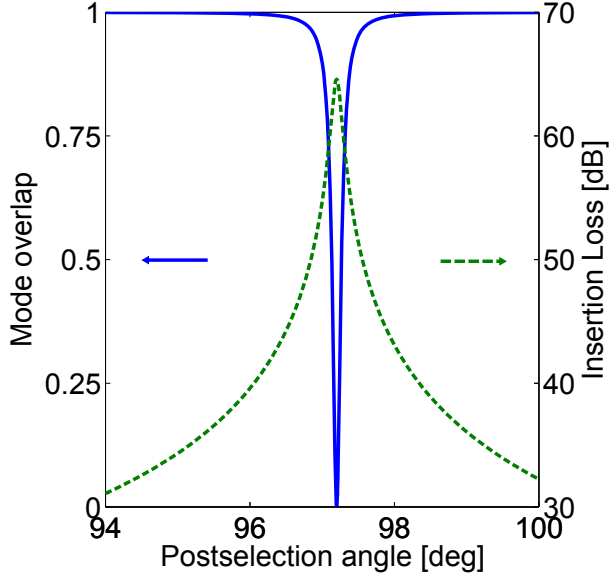}
\caption{Mode overlap and insertion loss as a function of the post-selection angle. Mode overlap $\rho$
of the mode functions corresponding to the quantum states with $\tau=0$ and
$\tau=100$ as, as a function of the post-selection angle $\theta$ (solid blue line).
The insertion loss, given by  $10\log_{10}\,P_{\mathrm{out}}/P_{\mathrm{in}}$ is indicated by the dotted green line. The minimum mode overlap, and maximum insertion loss, corresponds to the post-selection angle
$\theta$ that fulfils the condition $\omega_0 \tau-\Gamma=\pi$, which corresponds to $\theta=97.2^{\circ}$. Data: $\lambda_0=1.5 \,\mu$m, $T_0=100$ fs.}
\label{figure_overlap_loss}
\end{figure}

In the WVA scheme considered here, the signal after PBS$_2$ is
projected into the orthogonal polarisation states $\hat{u}$ and
$\hat{v}$, and as a result the signals in both output ports are
given by Eqs. (\ref{projections1}) and (\ref{projections2}).
Making use of Eqs. (\ref{projections1}), (\ref{projections2}) and
(\ref{overlap2}) one obtains that the mode overlap (for $\Phi_u$)
reads
\begin{equation}
\rho =\frac{1+\cos \Gamma+\gamma \cos
\omega_0 \tau+\gamma \cos (\omega_0 \tau-\Gamma) - i \left[ \sin
\Gamma+\gamma \sin \omega_0 \tau+\gamma \sin (\omega_0
\tau-\Gamma) \right]}{2\left[1+\cos \Gamma \right]^{1/2}
\left[1+\gamma \cos (\omega_0 \tau-\Gamma) \right]^{1/2}}.
\end{equation}
For $\tau=0$, and therefore $\gamma=1$, we obtain $\rho=1$. Fig.
3 shows the mode overlap of the signal in
the corresponding output port for a delay of $\tau=100$ as. The
mode overlap has a minimum for $\omega_0 \tau-\Gamma=\pi$, where
the two mode functions becomes easily distinguishable, as shown in
Fig. 2 (a). The effect of the polarisation
projection, a key ingredient of the WVA scheme, can be understood
as a change of the mode overlap ({\em mode distinguishability})
between states with different delay $\tau$.

However, an enhanced mode distinguishability in this output port
is accompanied by a corresponding increase of the insertion loss,
as it can be seen in Fig. 3. The insertion
loss, $P_{\mathrm{out}}(\tau)/P_{\mathrm{in}}=1/2\, \left[
1+\gamma \cos (\omega_0 \tau-\Gamma)\right]$, is the largest when
the modes are close to orthogonal ($\rho \sim 0$). Both effects
indeed compensate, as it should be, since WVA implements unitary
transformations, and the trace distance between quantum states is
preserved under unitary transformations. The quantum overlap
between the states reads
\begin{eqnarray}
& & |\langle \Phi_u(\tau_0)|\Phi_u(\tau_0+\tau)\rangle|^2=|\langle
\Phi_v(\tau_0)|\Phi_v(\tau_0+\tau)\rangle|^2 \nonumber \\
& & =\exp \left[-\frac{N}{2} \left( 1-\gamma \cos \omega_0
\tau\right) \right],
\end{eqnarray}
so
\begin{eqnarray}
& & |\langle \Phi_0|\Phi_1 \rangle|^2 \nonumber \\
& & =|\langle \Phi_u(\tau_0)|\Phi_u(\tau_0+\tau)\rangle_u|^2
|\langle \Phi_v(\tau_0)|\Phi_v(\tau_0+\tau)\rangle_v|^2 \nonumber \\
& & =\exp \left[-N \left( 1-\gamma \cos \omega_0 \tau\right)
\right] \label{output_result},
\end{eqnarray}
which is the same result [see Eq. (\ref{input_result})] obtained
for the signal after PBS$_1$, but before PBS$_2$.

We can also see the previous results from a slightly different
perspective making use of the Cram\'{e}r-Rao inequality
\cite{helstrom1976}. The WVA scheme considered throughout can be
thought as a way of estimating the value of the single parameter
$\tau$ with the help of a light pulse in a coherent state $|\alpha
\rangle$. Since the quantum state is pure, the minimum variance
that can show any unbiased estimation of the parameter $\tau$, the
Cram\'{e}r-Rao inequality, reads
\begin{equation}
\mathrm{Var} \left( \hat{\tau} \right) \ge \frac{1}{4}\left[ \langle
\frac{\partial \alpha}{\partial \tau}| \frac{\partial
\alpha}{\partial \tau} \rangle-\left| \langle \alpha |
\frac{\partial \alpha}{\partial \tau}\rangle\right|^2
\right]^{-1}, \label{cramer1}
\end{equation}
Making use of Eq. (\ref{state1}), one obtains that here the
Cram\'{e}r-Rao inequality reads \cite{derivative}
\begin{equation}
\mathrm{Var} \left( \hat{\tau} \right) \ge \frac{1}{2N \left(
\omega_0^2+B^2\right)} \label{cramer2}
\end{equation}
where $B=\sqrt{2 \ln 2}/T_0$ is the rms bandwidth in angular
frequency of the pulse. In all cases of interest $B \ll
\omega_0$. The Cram\'{e}r-Rao inequality is a fundamental limit
that set a bound to the minimum variance that any measurement can
achieve. It is unchanged  by unitary transformations and only
depends on the quantum state considered.

Inspection of Eqs. (\ref{input_result}) and (\ref{output_result})
seems to indicate that a measurement after projection in any
basis, the core element of the weak amplification scheme, provides
no fundamental metrological advantage. Notice that this result
implies that the only relevant factor limiting the sensitivity of
detection is the quantum nature of the light used (a {\em coherent
state} in our case). To obtain this result, we are implicitly
assuming that a) we have full access to all relevant
characteristics of the output signals; and b) detectors are ideal,
and can detect any change, as small as it might be, if enough
signal power is used. If this is the case, weak value
amplification provides no enhancement of the sensitivity.

However, this can be far from truth in many realistic experimental
situations. In the laboratory, the quantum nature of light is an
important factor, but not the only one, limiting the capacity to
measure tiny changes of variables of interest. On the one hand,
most of the times we detect only certain characteristic of the
output signals, probably the most relevant, but this is still partial information
about the quantum state. On the other hand, detectors are not ideal and
noteworthy limitations to its performance can appear. To name a
few, they might no longer work properly above a certain photon
number input, electronics and signal processing of data can limit
the resolution beyond what is allowed by the specific quantum
nature of light, conditions in the laboratory can change randomly
effectively reducing the sensitivity achievable in the experiment.
Surely, all of these are {\em technical} rather than {\em
fundamental} limitations, but in many situations the ultimate
limit might be {\em technical} rather than {\em fundamental}. In
this scenario, we show below that weak value amplification can be
a {\em valuable} and an {\em easy} option to overcome all of these
technical limitations, as it has been demonstrated in numerous
experiments.

\begin{figure}
\centering
\includegraphics[width=0.9\textwidth]{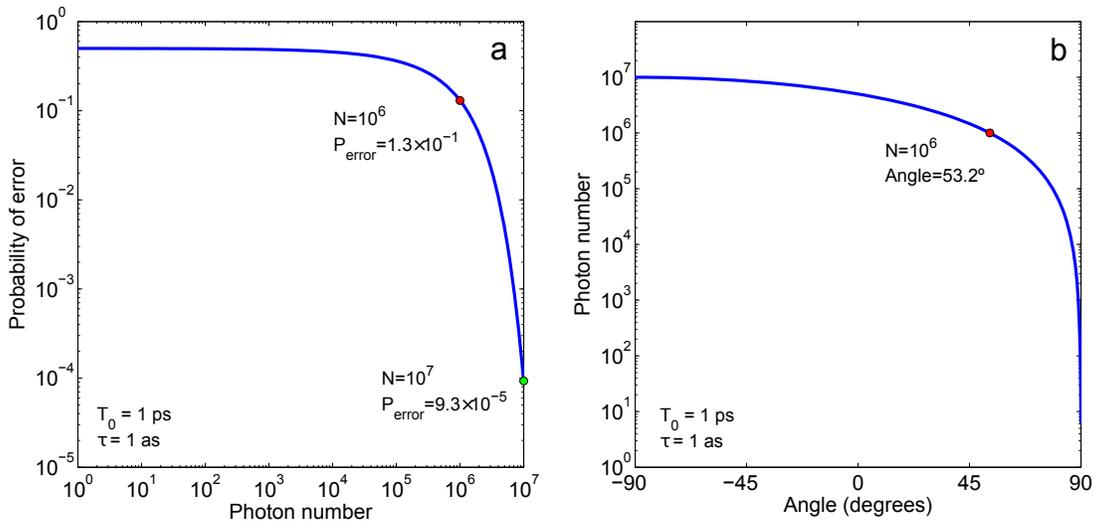}
\caption{Reduction of the probability of error using a weak value amplification scheme.
(a) Minimum probability of error as a function of the
photon number $N$ that leaves the interferometer. The two points
highlighted corresponds to $N=10^6$, which yields
$P_{\mathrm{error}}=1.3 \times 10^{-1}$, and $N=10^7$, which
yields $P_{\mathrm{error}}=9.3 \times 10^{-5}$. (b) Number of
photons ($N_{\mathrm{out}}$) after projection in the polarisation
state $\hat{u}=1/\sqrt{2} \left[ \hat{x}+\hat{y}\exp(i\theta)
\right]$, as a function of the angle $\theta$. The input number of
photons is $N=10^7$. The dot corresponds to the point
$N_{\mathrm{out}}=10^6$ and $\theta=53.2^{\circ}$. Pulse width:
$T_0$=1 ps; temporal delay: $\tau$= 1 as.} \label{figure_sat}
\end{figure}

\section*{Discussion}
\subsection*{Advantages of using weak value amplification (I): when the detector cannot work above a certain photon number.}
Let us suppose that we have at hand light detectors that cannot be
used with more than $N_0$ photons. Any limitation on the detection
time or the signal power would produce such limitation. The
technical advantages of using WVA in this scenario has been
previously pointed out \cite{jordan2014}. Here we make this
apparent from a quantum estimation point of view, and quantify
this advantage.

Fig. 4(a) shows the minimum probability of error as
a function of the number of photons ($N$) entering (and leaving)
the interferometer. For $N_0=10^6$, inspection of the figure shows
that the probability of error is $P_\mathrm{error}=1.3 \times
10^{-1}$. This is the best we can do with this experimental scheme
and these particular detectors without resorting to weak value
amplification. However, if we project the output signal from the
interferometer into a specific polarisation state, and increase
the flux of photons, we can decrease the probability of error,
without necessarily going to a regime of high depletion of the
signal \cite{torres2012}. For instance, with
$\theta=53.2^{\circ}$, and a flux of photons of $N=10^7$, so that
after projection $N_{\mathrm{out}}=10^6$ photons reach the
detector, the probability of error is decreased to
$P_{\mathrm{error}}=9.3 \times 10^{-5}$, effectively enhancing the
sensitivity of the experimental scheme (see Fig. 4(b)). The probability of error can be further
decreased, also for other projections, at the expense of further
increasing the input signal $N$.

In general, the minimum quantum overlap achievable between the
states without any projection is
\begin{equation}
|\langle \Phi_0|\Phi_1\rangle|^2=\exp \left[-N_0 \left( 1-\gamma
\cos \omega_0 \tau\right) \right],
\end{equation}
while making use of projection in a weak value amplification
scheme is
\begin{equation} |\langle \Phi_0|\Phi_1\rangle|^2=\exp
\left[-\frac{2N_0 \left(1-\gamma \cos \omega_0 \tau
\right)}{1+\gamma \cos \left(\omega_0 \tau-\Gamma-\pi/2 \right)}
\right]. \label{enhancement}
\end{equation}
Eq. (\ref{enhancement}) shows that when the number of photons that
the detection scheme can handle is limited ($N_0$), projection
into a particular polarisation state, at the expense of increasing
the signal level, is advantageous. From a quantum estimation point
of view, WVA increases the minimum probability of error reachable,
since the projection makes possible to use the maximum number of
photons available ($N_0$) with a corresponding enhanced mode
overlap. Notice that the effect of using different polarisation
projections can be beautifully understood as reshaping of the
balance between signal level and mode overlap.

\subsection*{Advantages of using weak value amplification (II): when the detector cannot differentiate between two signals}
As second example, let us consider that specific experimental
conditions makes hard, even impossible, to detect very similar
modes, i.e., with mode overlap $\rho \sim 1$. We can represent
this by assuming that there is an {\em effective} mode overlap
($\rho_{\mathrm{eff}}$) which takes into account all relevant
experimental limitations of a specific set-up, given by
\begin{equation}
\rho \Longrightarrow \rho_{\mathrm{eff}}=1-(1-\rho)\exp \left[
-\left(\frac{\rho}{a}\right)^n \right].
\end{equation}
Fig. 5 shows an example where we assume
that detected signals corresponding to $\rho > 0.9$ cannot be
safely distinguished due to technical restrictions of the
detection system. For $\rho > 0.9$, $\rho_{\mathrm{eff}}=1$, so
the detection system cannot distinguish the states of interest
even by increasing the level of the signal. On the contrary, for
smaller values of $\rho$, accessible making use of a weak
amplification scheme, this limitation does not exist since the
detection system can resolve this modes when enough signal is
present.

\begin{figure}
\centering
\includegraphics[width=0.5\textwidth]{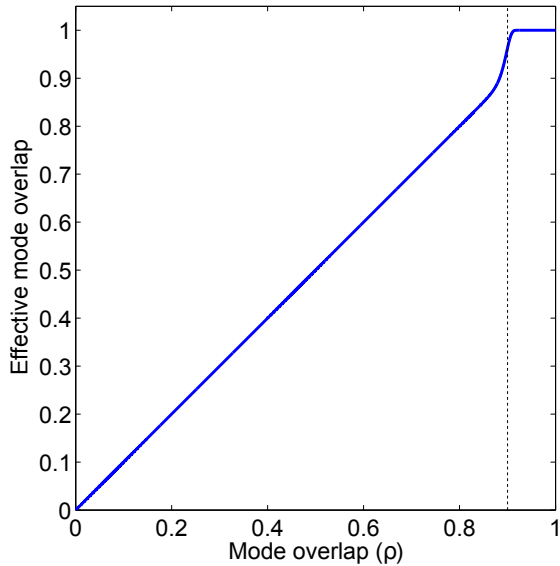}
\caption{Effective mode overlap. For $\rho>0.9$ the detection system cannot distinguish the states of interest. Data: $a=0.9$ and $n=100$.} \label{figure_effective_rho}
\end{figure}

\subsection*{Advantages of using weak value amplification (III): enhancement of the Fisher information}
Up to now, we have used the concept of trace distance to look for
the minimum probability of error achievable in {\em any}
measurement when using a given quantum state. In doing that, we
only considered how the quantum state changes for different values
of the variable to be measured, without any consideration of how
this quantum state is going to be detected.  If we would like to
include in the analysis additional characteristics of the
detection scheme, one can use the concept of Fisher information,
that requires to consider the probability distribution of possible
experimental outcomes for a given value of the variable of
interest. In this approach, one chooses different probability
distributions to describe formally {\em characteristics} of
specific detection scheme \cite{jordan2014}.

Let us assume that to estimate the value of the delay $\tau$, we
measure the shift of the centroid ($\Delta f$) of the spectrum
$\Phi_u(\tau)$, given by Eq. (\ref{projections2}). A particular
detection scheme will obtain a set of results $\left\{ (\Delta
f)_i \right\}$, $i=1..N$ for a given delay $\tau$. $N$ is the
number of photons detected. The Fisher information $I(\tau)$
provides a bound of $\mathrm{Var}\left( \hat{\tau} \right)$ for
any unbiased estimator when the probability distribution
$p(\left\{ (\Delta f)_i \right\}|\tau)$ of obtaining the set
$\left\{ (\Delta f)_i \right\}$, for a given $\tau$, is known. If
we assume that the probability distribution $p(\left\{(\Delta f)_i
\right\} |\tau)$ is Gaussian, with mean value $\Delta f$ given by
Eq. (\ref{centroid_shift}) and variance $\sigma^2$, determined by
the errors inherent to the detection process, the Fisher
information reads \cite{fisher1}
\begin{equation}
I(\tau)=\frac{N}{\sigma^2} \left[ \frac{\partial \Delta
f}{\partial \tau}\right]^2
\end{equation}
where
\begin{equation}
\frac{\partial \Delta f}{\partial \tau}=\frac{\gamma B^2 \left[
B^2 \tau^2 \sin \phi-\omega_0 \tau \left( \gamma+\cos
\phi\right)-\sin \phi \left( 1+\gamma \cos \phi\right)
\right]}{2\pi\, \left( 1+\gamma \cos \phi\right)^2}
\end{equation}
and $\phi=\omega_0 \tau-\Gamma$.

For $\phi=0$, i.e., the angle of post-selection is
$\theta=-\pi/2+\omega_0 \tau$, the Fisher information is
\begin{equation}
I_0=\frac{N_0}{2} \left( 1+\gamma\right) \times \frac{\gamma^2 B^4
(\omega_0 \tau)^2}{2\pi\sigma^2 (1+\gamma)^2}= \frac{\gamma^2 B^4
(\omega_0 \tau)^2}{4\pi\sigma^2 (1+\gamma)} \label{fisher_bound1}
\end{equation}

Notice that $\theta=-\pi/2$ corresponds to considering equal input
and output polarization state, i.e., no weak value amplification
scheme. For $\phi=\pi$, where the angle of post-selection is
$\theta=\pi/2+\omega_0 \tau$, we have
\begin{equation}
I_{\pi}=\frac{N_0}{2} \left( 1-\gamma\right) \times \frac{\gamma^2
B^4 (\omega_0 \tau)^2}{2\pi\sigma^2 (1-\gamma)^2}= \frac{\gamma^2
B^4 (\omega_0 \tau)^2}{4\pi\sigma^2 (1-\gamma)}
\label{fisher_bound2}
\end{equation}
$\theta=\pi/2$ corresponds to considering an output polarisation
state orthogonal to the input polarisation state i.e., when the
effect of weak value amplification is most dramatic, as it can be
easily observed in Fig. 2(a). The Fisher bound for $\Phi=\pi$
is a factor $I_{\pi}/I_0=(1+\gamma)/(1-\gamma)$ larger than the
bound for $\Phi=0$, so WVA achieves enhancement of the Fisher
information. This Fisher information enhancement effect, which
does not happen always, it has been observed for certain WVA
schemes \cite{viza2013,jordan2014}.

There is no contradiction between the facts that the minimum
probability of error, obtained by making use of the concept of
trace distance, is not changed by WVA, while at the same time
there can be enhancement of the Fisher information. By selecting a
particular probability distribution to evaluate the Fisher
information, we include information about the detection scheme. In
our case, we estimate the value of $\tau$ by measuring the
$\tau$-dependent shift of the centroid of the spectrum of the
signal in one output port after PBS$_2$, which is only part of all
the information available, given by the full signal in Eqs.
(\ref{projections1}) and (\ref{projections2}). We also assumed a
Gaussian probability distribution with a constant variance
$\sigma^2$ independent of $\tau$. The Cram\'{e}r-Rao bound we have
derived here depends on the full information available (the
quantum state) before any particular detection. An unitary
transformation, as WVA is, does not modify the bound. On the
contrary, the Fisher information, by using a particular
probability distribution to describe the possible outcomes in an
particular experiment, selects certain aspects of the
quantum state to be measured ({\em partial information}), and this
bound can change in a WVA scheme, although the bound should
be always above the Cram\'{e}r-Rao bound. In this
restrictive scenario, the use of certain polarization projections
can be preferable.

The existence and nature of these different bounds might possibly
explain certain confusion about the capabilities of WVA, whether
WVA is considered to provide any metrological advantage or not. On
the one hand, if we consider the trace distance, or the quantum
Cram\'{e}r-Rao inequality, without any consideration about how the
quantum states are detected, post-selection inherent in WVA does
not lower the minimum probability of error achievable, so from
this point of view WVA offers no metrological advantage. On the
other hand, in certain scenarios, the Fisher information, when it
takes into account {\em information about the detection scheme},
can be enhanced due to post-selection. In this sense, one can
think of WVA as an advantageous way to optimize a particular
detection scheme.

\section*{Conclusions}
WVA schemes makes use of linear optics unitary transformations.
Therefore, if the only limitations in a measurement are due to the
quantum nature ({\em intrinsic statistics}) of the light, for
instance, the presence of Shot noise in the case of coherent
beams, WVA does not offer any advantage regarding any decrease of
the minimum probability of error achievable. This is shown by
making use of the trace distance between quantum states or the
Cram\'{e}r-Rao inequality, which set sensitivity bounds that are
independent of any particular post-selection. However, notice that
this implicitly assume that full information about the quantum
states used can be made available, and detectors are ideal, so
they can detect any change of the variable of interest, as small
as it might be, provided there is enough signal power.

Nevertheless, these assumptions are in many situations of interest
far from true. These limitations, sometimes refereed as {\em
technical noise}, even though not fundamental (one can always
imagine using a better detector or a different detection scheme)
are nonetheless important, since they limit the accuracy of
specific detection systems at hand. In these scenarios, the
importance of weak value amplification is that by decreasing the
mode overlap associated with the states to be measured and
possibly increasing the intensity of the signal, the weak value
amplification scheme allows, in principle, to distinguish them
with lower probability of error.

We have explored some of these scenarios from an quantum
estimation theory point of view. For instance, we have seen that
when the number of photons usable in the measurement is limited,
the minimum probability of error achievable can be effectively
decreased with weak value amplification. We have also analyzed how
weak value amplification can differentiate between {\em in
practice}-indistinguishable states by decreasing the mode overlap
between its corresponding mode functions.

Finally we have discussed how the confusion about the usefulness
of weak value amplification can possibly derive from considering
different bounds related to how much sensitivity can, in
principle, be achieved when estimating a certain variable of
interest. One might possibly say that the advantages of WVA
{\em have nothing to do with fundamental limits and should not be
viewed as addressing fundamental questions of quantum mechanics}
\cite{caves2014}. However, {\em from a practical rather than
fundamental point of view}, the use of WVA can be advantageous in
experiments where sensitivity is limited by experimental
(technical), rather than fundamental, uncertainties. In any case,
if a certain measurement is {\em optimum} depends on its
capability to effectively reach any bound that might exist.

\newpage



\newpage
\section*{Figure Captions}
\subsection*{Figure1}
Weak value amplification scheme aimed at detecting
extremely small temporal delays. The input pulse polarisation state is selected to
be left-circular by using a polariser, a quarter-wave plate
(QWP) and a half-wave plate (HWP). A first polarising beam splitter
(PBS$_1$) splits the input into two orthogonal linear polarisations
that propagate along different arms of the interferometer. An
additional QWP is introduced in each arm to rotate the beam
polarisation by $90^{\circ}$ to allow the recombination of both beams,
delayed by a temporal delay $\tau$, in a single beam by the same PBS. After PBS$_1$, the output
polarisation state is selected with a liquid crystal variable
retarder (LCVR) followed by a second polarising beam splitter (PBS$_2$). The variable
retarder is used to set the parameter $\theta$ experimentally.
Finally, the spectrum of each output beam is measured using an optical spectrum analyzer (OSA).
($\hat{x}$,$\hat{y}$) and ($\hat{u}$,$\hat{v}$) correspond to two sets of
orthogonal polarisations.

\subsection*{Figure2}
Spectrum measured at the output. (a) and (b): Spectral shape of the mode functions for $\tau=0$  (solid blue line) and $\tau=100$ as (dashed green line). In (a) the post-selection angle $\theta$ is $97.2^{\circ}$, so as to fulfil the condition $\omega_0 \tau-\Gamma=\pi$. In (b) the angle $\theta$ is $96.7^{\circ}$. (c) Shift of the centroid of the spectrum of the output pulse after projection into the polarisation state $\hat{u}$ in PBS$_2$, as a function of the post-selection angle $\theta$. Green solid line: $\tau=10$ as; Dotted red line: $\tau=50$ as, and dashed blue line: $\tau=100$ as. Label {\bf I} corresponds to $\theta=96.7^{\circ}$ [mode for $\tau=100$ as shown in (b)]. Label {\bf II} corresponds to $\theta=97.2^{\circ}$, where the condition $\omega_0 \tau-\Gamma=\pi$ is fulfiled [mode for $\tau=100$ shown in (a)]. It yields the minimum mode overlap between states with $\tau=0$ and $\tau \neq 0$. Data: $\lambda_0=1.5\, \mu$m and $T_0=100$ fs.

\subsection*{Figure3}
Mode overlap and insertion loss as a function of the post-selection angle. Mode overlap $\rho$
of the mode functions corresponding to the quantum states with $\tau=0$ and
$\tau=100$ as, as a function of the post-selection angle $\theta$ (solid blue line).
The insertion loss, given by  $10\log_{10}\,P_{\mathrm{out}}/P_{\mathrm{in}}$ is indicated by the dotted green line. The minimum mode overlap, and maximum insertion loss, corresponds to the post-selection angle
$\theta$ that fulfils the condition $\omega_0 \tau-\Gamma=\pi$, which corresponds to $\theta=97.2^{\circ}$. Data: $\lambda_0=1.5 \,\mu$m, $T_0=100$ fs.

\subsection*{Figure4}
Reduction of the probability of error using a weak value amplification scheme.
(a) Minimum probability of error as a function of the
photon number $N$ that leaves the interferometer. The two points
highlighted corresponds to $N=10^6$, which yields
$P_{\mathrm{error}}=1.3 \times 10^{-1}$, and $N=10^7$, which
yields $P_{\mathrm{error}}=9.3 \times 10^{-5}$. (b) Number of
photons ($N_{\mathrm{out}}$) after projection in the polarisation
state $\hat{u}=1/\sqrt{2} \left[ \hat{x}+\hat{y}\exp(i\theta)
\right]$, as a function of the angle $\theta$. The input number of
photons is $N=10^7$. The dot corresponds to the point
$N_{\mathrm{out}}=10^6$ and $\theta=53.2^{\circ}$. Pulse width:
$T_0$=1 ps; temporal delay: $\tau$= 1 as.

\subsection*{Figure5}
Effective mode overlap. For $\rho>0.9$ the detection system cannot distinguish the states of interest. Data: $a=0.9$ and $n=100$.

\newpage

\small{\textbf{Acknowledgements} We acknowledge support from the
Severo Ochoa program (Government of Spain), from the ICREA
Academia program (ICREA, Generalitat de Catalunya) and from
Fundaci\'{o} Privada Cellex, Barcelona.}

\small{\textbf{Author Contributions} All authors contribute
equally to the paper and reviewed the manuscript.}

\small{\textbf{Additional information} Competing financial
interests: The authors declare no competing financial interests.}


\begin{thebibliography}{19}
\bibitem{derivative} Let us define $Q=\langle\frac{\partial \alpha}{\partial \tau}| \frac{\partial
\alpha}{\partial \tau} \rangle-\left| \langle {\small \alpha} | \frac{\partial \alpha}{\partial \tau} \rangle\right|^2$. For a coherent product state of the form $|\alpha\rangle=\prod_i |\alpha_i \rangle$, where the index $i$ refers to different frequency modes, one obtains that $Q=\sum_i Q_i$, where $Q_i=\langle \frac{\partial \alpha_i}{\partial \tau}| \frac{\partial \alpha_i}{\partial \tau} \rangle-\left| \langle
{\small \alpha_i} | \frac{\partial \alpha_i}{\partial \tau} \rangle\right|^2$. If $\alpha_i=\beta_i \exp(i\varphi_i)$, where $|\beta_i|^2$ is the mean number of photons in frequency mode $i$ and only $\varphi_i$ depends on the parameter $\tau$ as $\varphi_i=(\omega_0+\Omega_i)\tau$, one obtains that $|\partial
\alpha_i /\partial \tau \rangle=i \left(\partial \varphi_i/\partial \tau\right) \alpha_i a_i^{\dagger}|\alpha_i
\rangle$, where $a_i^{\dagger}$ is the creation operator of the corresponding frequency mode.
\bibitem{fisher1} The unknown parameter of interest has value
$\tau$. After repeated measurements to estimate its value, we
obtain a distribution of outcomes $\left\{ x \right\}$ which can
be characterized by a probability distribution $p(x|\tau)$ that
depends on the value of $\tau$. The Fisher information reads
$I(\tau)=-\int dx p(x|\tau)\partial_{\tau}^2 p(x|\tau)$. The
variance of any unbiased estimator that makes use of the ensemble
$\left\{ x \right\}$ is bounded from below by $\mathrm{Var}(\hat{\tau})
\ge 1/I(\tau)$. When the Fisher function can be written as $I[\eta(\tau)]$,
where $\eta$ is the variable that we measure, the Fisher information can be written
as $I(\tau) = I(\eta)\left(\frac{\partial\eta}{\partial\tau}\right)^2$.
\end{thebibliography}
\end{document}